\documentclass[aps,preprint,superscriptaddress,prx]{revtex4-2}

\usepackage{graphicx}
\usepackage{dcolumn}
\usepackage{bm}
\usepackage{xcolor}
\usepackage{amssymb}   
\usepackage{amsmath}

\begin{document}

\title{
Thermodynamic evidence for field-angle dependent Majorana gap in a Kitaev spin liquid
}

\date{\today}

\author{O.~Tanaka}
\author{Y.~Mizukami} \email{mizukami@edu.k.u-tokyo.ac.jp}
\author{R.~Harasawa}
\author{K.~Hashimoto}
\affiliation{Department of Advanced Materials Science, University of Tokyo, Kashiwa, Chiba 277-8561, Japan}
\author{N.~Kurita}
\author{H.~Tanaka}
\affiliation{Department of Physics, Tokyo Institute of Technology, Meguro, Tokyo 152-8551, Japan}
\author{S.~Fujimoto}
\affiliation{Department of Materials Engineering Science, Osaka University, Toyonaka, Osaka 560-8531, Japan}
\author{Y.~Matsuda}
\affiliation{Department of Physics, Kyoto University, Sakyo-ku, Kyoto 606-8502, Japan}
\author{E.-G.~Moon}
\affiliation{Department of Physics, Korea Advanced Institute of Science and Technology (KAIST), Daejeon 34141, Korea}
\author{T.~Shibauchi} \email{shibauchi@k.u-tokyo.ac.jp}
\affiliation{Department of Advanced Materials Science, University of Tokyo, Kashiwa, Chiba 277-8561, Japan}

\maketitle

{\bf 
The exactly-solvable Kitaev model of two-dimensional honeycomb magnet leads to a quantum spin liquid (QSL) characterized by Majorana fermions, relevant for fault-tolerant topological quantum computations.
In the high-field paramagnetic state of $\bm{\alpha}$-RuCl$_{\bm{3}}$, half-integer quantization of thermal Hall conductivity has been reported as a signature of edge current, but the bulk nature of this state remains elusive.
Here, from high-resolution heat capacity measurements under in-plane field rotation, we find strongly angle-dependent low-energy excitations in bulk $\bm{\alpha}$-RuCl$_{\bm{3}}$.  
The excitation gap has a sextuple node structure, and the gap amplitude increases with field, exactly as expected for itinerant Majorana fermions in the Kitaev model.
Our thermodynamic results are fully linked with the transport quantization properties, providing the first demonstration of the bulk-edge correspondence in a Kitaev QSL.
Moreover, at higher fields where the quantum thermal Hall effect vanishes, we find the possible emergence of a novel nematic QSL state with two-fold rotational symmetry.
}


\newpage

Quantum spin liquids (QSLs) are enigmatic states of matter, in which quantum fluctuations and frustrations prevent spin configurations in a lattice from any solid-like ordered alignments~\cite{Anderson1973,Savary2016}. 
In the exactly solvable model of two-dimensional honeycomb lattice proposed by Kitaev~\cite{Kitaev2006}, the bond-dependent Ising interactions act as an exchange frustration, leading to a QSL ground state with characteristic excitations of Majorana fermions. 
These Majorana excitations are important to make non-abelian anyons that are useful for fault-tolerant topological quantum computations. 
Realizing this intriguing QSL state in real materials is therefore quite important, and there are tremendous efforts to search the QSL states in Mott insulators with strong spin-orbit coupling~\cite{Jackeli2009,Motome2020,Takagi2019}.

\begin{figure}[h]
    \centering
    \includegraphics[width=0.9\linewidth,pagebox=artbox]{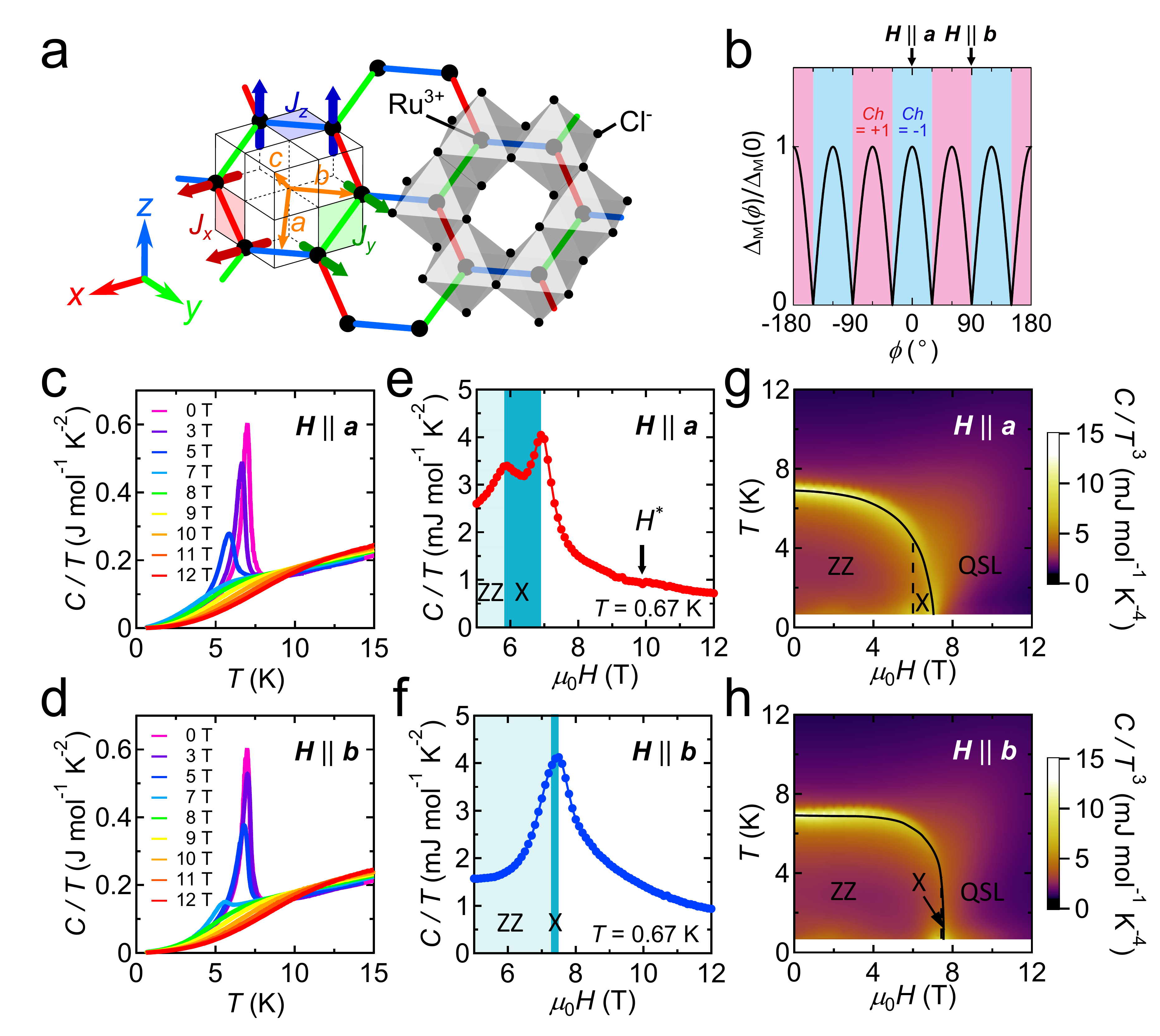}
    \caption{{\bf Field-temperature phase diagram of $\bm{\alpha}$-RuCl$_{\bm{3}}$.} {\bf a}, Schematic crystal structure of $\alpha$-RuCl$_3$ and definitions of the crystallographic axes ($a, b, c$) and the Ising spin axes ($x, y, z$). 
    Owing to the bond-dependent Ising interactions ($J_x, J_y, J_z$), each spin axis is perpendicular to the plane including a Ru-Ru bond and a shared edge of Cl octahedrons (colored squares). 
    Magnetic field $\bm{H}$ is applied within the honeycomb plane, and $\phi$ is the angle between $\bm{H}$ and the $a$ axis.
    {\bf b}, Majorana gap magnitude $\Delta_{\rm M}$ in Eq.\,(\ref{Deq}) as a function of in-plane field angle in the Kitaev model, which follows $|\cos{3\phi}|$ dependence. The Chern number is defined by the sign of $h_xh_yh_z$, and is zero for $\bm{H} \parallel \bm{b}$ and equivalent directions. 
    {\bf c}, {\bf d}, Temperature dependence of specific heat divided by temperature, $C/T$, at several fields for $\bm{H} \parallel \bm{a}$ ({\bf c}) and $\bm{H} \parallel \bm{b}$ ({\bf d}).
    {\bf e}, {\bf f}, Field dependence of $C/T$ at $T=0.67$\,K for $\bm{H} \parallel \bm{a}$ ({\bf e}) and $\bm{H} \parallel \bm{b}$ ({\bf f}). The arrow represents an $H^\ast$ anomaly (open square in Fig.\,\ref{Cangle}{\bf a}).
    {\bf g}, {\bf h}, $T$-$H$ phase diagrams of $\alpha$-RuCl$_3$ for $\bm{H} \parallel \bm{a}$ ({\bf g}) and $\bm{H} \parallel \bm{b}$ ({\bf h}). Superimposed color map represents the magnitude of $C/T^3$. Solid (dashed) line represents the phase boundary between zig-zag (ZZ) magnetic and paramagnetic QSL (intermediate X) phases.
    }
    \label{Hdiagram}
\end{figure}

In the Kitaev model~\cite{Kitaev2006}, each $S=1/2$ spin at the honeycomb site can be converted to two kinds of Majorana fermions, itinerant and localized ones, the latter of which form the so-called $Z_2$ flux (which may also be called as vison) per hexagon plaquette. 
By this representation the quantum many-body problem of spins can be simplified as one-body physics of moving Majorana fermions on localized $Z_2$ fluxes, and the ground state is found to be a QSL state with no long-range order. 
As the $Z_2$ flux has a sizeable gap $\Delta_{\rm flux}$, the low-energy behaviors are governed by the itinerant Majorana excitations~\cite{Nasu2015}. 
When one of the bond-dependent Ising interactions ($J_x$, $J_y$, or $J_z$) is larger than the sum of the other two, $|J_i|>|J_j|+|J_k|$ $(i,j,k=x,y,z)$, the excitations are gapped (A phases, or toric code phases). Otherwise the itinerant Majorana fermions at zero field exhibit gapless excitations described by cone-shaped dispersions with nodal points at zero energy (B phase), similar to the Dirac-type electronic structure of graphene which also has a two-dimensional honeycomb structure.
In the B phase, the application of magnetic field changes the low-energy gapless linear dispersion of Majorana fermions to a gapped one, and the Majorana excitation gap $\Delta_{\rm M}$ is given by   
\begin{equation}
    \Delta_{\rm M}\propto\frac{|h_xh_yh_z|}{\Delta_{\rm flux}^2}.
    \label{Deq}
\end{equation} 
Here $h_x$, $h_y$, and $h_z$ are the $x$, $y$, and $z$ components of applied magnetic field $\bm{H}$, respectively, which are defined by the Ising spin axes (see Fig.\,\ref{Hdiagram}a). 
When the field-induced gap is finite, the Kitaev QSL has edge states, which are topologically protected by non-zero Chern number $Ch=h_xh_yh_z/|h_xh_yh_z|$ determined by the sign of the product $h_xh_yh_z$. 
Equation\,(\ref{Deq}) immediately indicates that the Majorana gap is strongly field-angle dependent, and has nodes at directions where the product $h_xh_yh_z$ is zero. 
Across the nodes, a topological phase transition between $Ch=-1$ and $Ch=+1$ states can be induced by a rotation of magnetic field. 
Therefore, the observations of field-angle dependent gap as well as gapless excitations at nodal directions can be taken as an experimental verification of the Kitaev QSL state with Majorana fermion excitations~\cite{Go2020}. 

In $\alpha$-RuCl$_3$, Ru$^{3+}$ ions which are surrounded by Cl$^{-}$ octahedrons form a layered honeycomb structure~\cite{Plumb2014}. 
As shown in Fig.\,\ref{Hdiagram}a, the crystallographic $a$ ($b$) direction is perpendicular (parallel) to the bond direction, which corresponds to $(1,1,-2)$ ($(-1,1,0)$) direction in the $(x,y,z)$ spin-axis coordinate. 
In this coordinate the product $h_xh_yh_z$ has the same rotational symmetry as $f_{xyz}$-wave, and thus if we rotate the magnetic field in the honeycomb plane that corresponds to $(111)$ plane, the Majorana gap changes as $\Delta_{\rm M}(\phi) \propto |\cos 3\phi|$, where $\phi$ is the angle between the in-plane field and $a$ axis (see Fig.\,\ref{Hdiagram}b). 

At zero field, $\alpha$-RuCl$_3$ exhibits an antiferromagnetic zig-zag (ZZ) order below $T_{\rm N}\approx 7$\,K~\cite{Johnson2015}, which indicates the presence of magnetic interactions of non-Kitaev type. 
The estimates from theoretical calculations and neutron scattering suggest that the ferromagnetic Kitaev interaction $J_{\rm K}$ (corresponding to B phase) and non-Kitaev interactions such as off-diagonal $\Gamma$ terms are dominant over small Heisenberg interactions~\cite{Winter2016,Do2017}. 
However, the peak in the temperature dependence of specific heat at $\sim100$\,K well above $T_{\rm N}$ is characterized by an entropy release of $(R/2)\ln2$ ($R$ is the gas constant)~\cite{Do2017,Widmann2019}, and Raman scattering spectra show fermionic responses~\cite{Sandilands2015,Nasu2016}, which have been explained by fractionalized excitations consistent with the Kitaev model. 
Several experimental studies have found that the application of magnetic field in the honeycomb plane can suppress the ZZ order with a critical field of $\sim 7$\,T~\cite{Yadav2016,Banerjee2018}, above which a paramagnetic state appears. 
Most notably, in a limited field ($H$) and temperature ($T$) range of this field-induced paramagnetic state, the field-dependent thermal Hall conductivity measurements~\cite{Kasahara2018,Yokoi2020preprint,Yamashita2020} have revealed a plateau behavior with the value close to one half of quantized thermal Hall conductivity of electronic system. 
This half-integer quantization is consistent with the presence of edge current of Majorana fermions that have a half degrees of freedom of electrons. 
However, the experimental difficulties of thermal Hall measurements at very low temperatures call for another approach to elucidating this intriguing state. 
Moreover, the nature of the QSL properties at low energies in the bulk is far from being understood, and thus the topological character of this state, especially one-to-one correspondence between the bulk and edge properties, remains to be verified. 
Here we focus on the heat capacity measurements under magnetic fields that are rotated within the honeycomb plane of $\alpha$-RuCl$_3$, from which low-energy excitations are investigated thermodynamically. 

The specific heat capacity $C$ of a high-quality single crystal of $\alpha$-RuCl$_3$ is measured down to $T\approx 0.6$\,K by a high-resolution long-relaxation technique with a piezo-based rotator in a 12-T superconducting magnet (see Supplementary Information). 
Figure\,\ref{Hdiagram}c,d shows the temperature dependence of $C/T$ measured under fixed fields applied parallel to the $a$ and $b$ axes, respectively. 
The antiferromagnetic transition is clearly identified by the peak anomaly in $C/T$ at $T_{\rm N}$, which is shifted to lower temperature with field. 
As shown in Fig.\,\ref{Hdiagram}e,f, the field dependence of $C/T$ for $\bm{H} \parallel \bm{a}$ and $\bm{H} \parallel \bm{b}$ at the lowest temperature exhibits two anomalies around 6--7.5\,T, which are consistent with the reported phase transitions from the ZZ phase to the intermediate so-called X phase and then to the high-field paramagnetic phase~\cite{Lampen2018,Balz2019}. 
The obtained $T$-$H$ phase diagrams for the two field directions are summarized in Fig.\,\ref{Hdiagram}g,h, superimposed on the magnitude of $C/T^3$. 
The $T_{\rm N}(H)$ curvature is more square-shaped and the critical fields are higher for $\bm{H} \parallel \bm{b}$. 

\begin{figure}[h]
    \centering
    \includegraphics[width=\linewidth,pagebox=artbox]{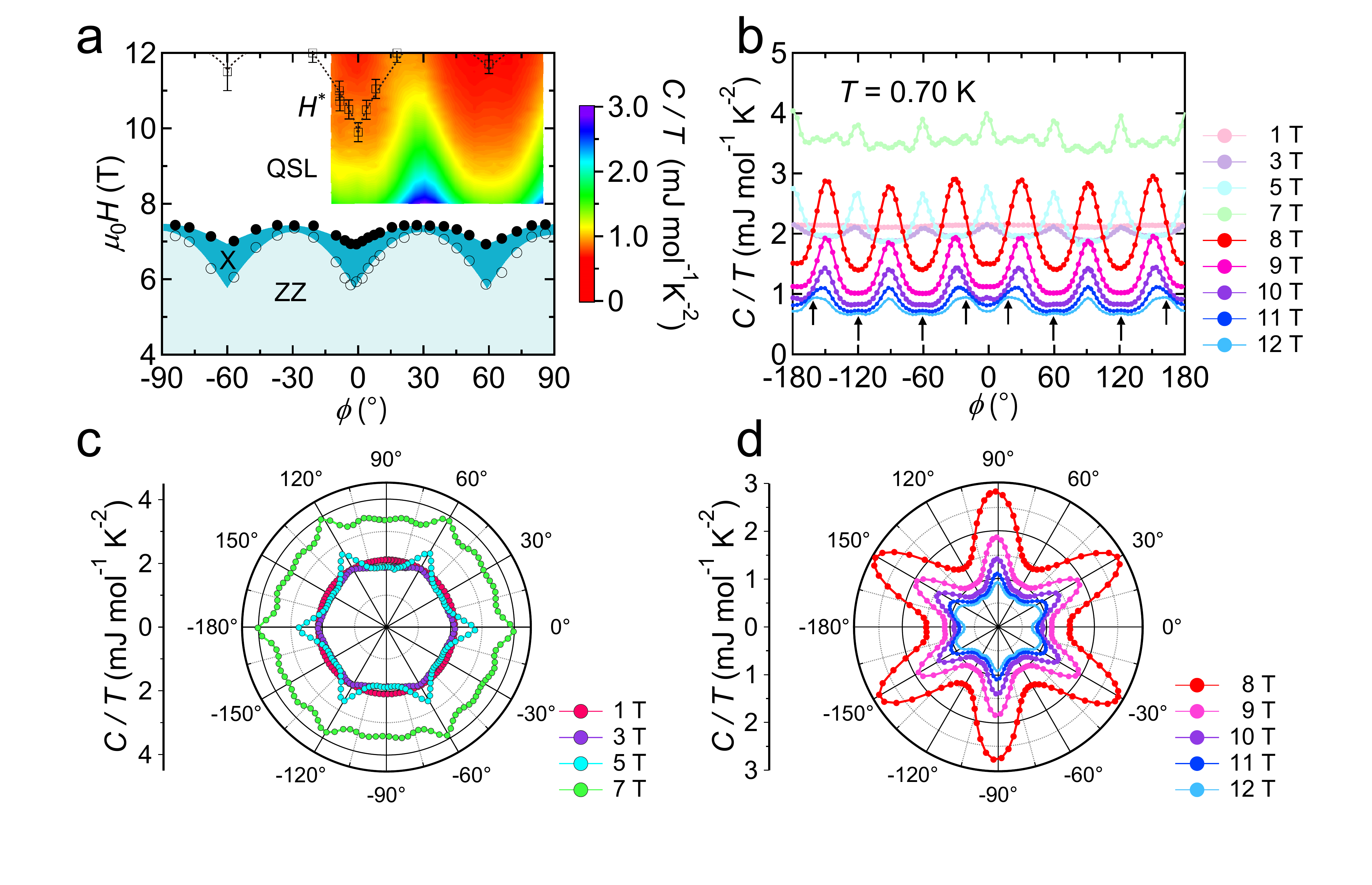}
    \caption{{\bf Field-angle dependence of specific heat.}
    {\bf a}, Field-angle phase diagram determined by the specific heat anomalies at $T=0.70$\,K. Open circles represent the transition from the zig-zag (ZZ) phase to X magnetic phase. Closed circles represent the transition between the antiferromagnetic and paramagnetic QSL phases. In the QSL state, the $H^\ast$ anomalies determined by field and angle dependence of specific heat (Fig.\,S4) are also shown (arrows in {\bf b}). Superimposed color map represents the magnitude of $C/T$.
    {\bf b}, Angle dependence of $C/T$ at fixed field magnitudes at $T=0.70$\,K. No data are shifted.
    {\bf c}, Polar plot of $C/T(\phi)$ below 7\,T. 
    {\bf d}, Polar plot of $C/T(\phi)$ above 8\,T in the paramagnetic QSL phase.    
    }
    \label{Cangle}   
\end{figure}

The phase boundaries of the ZZ, intermediate X, and high-field paramagnetic phases determined by the specific heat anomalies are shown in the $H$-$\phi$ diagram of Fig.\,\ref{Cangle}a. 
The critical field that separates antiferromagnetic and paramagnetic phases falls within a range between 6.9 and 7.4\,T, showing a kink behavior associated with the spin flop effect due to field rotation~\cite{Lampen2018}. 
Figure\,\ref{Cangle}b represents the angular dependence of $C/T$ at $T=0.70$\,K for several fixed field magnitudes. 
At low fields in the antiferromagnetic state, the specific heat has maxima along the $a$ axis and equivalent angles ($\phi=n\times 60^{\circ}$, where $n$ is integer). 
In stark contrast, it has maxima along the $b$ axis and equivalent angles ($\phi=n\times 60^{\circ}+ 30^{\circ}$) in the high-field paramagnetic state. 
This remarkable phase reversal between the two different states can also be clearly seen in the contrasting polar plots in Fig.\,\ref{Cangle}c,d. 
The angular dependence below $\sim 5$\,T in the ZZ phase has a kink behavior at the maxima, which is probably related to the spin flop effect~\cite{Go2020}. 
In contrast, the angle dependence in the paramagnetic phase is much rounded, implying a different mechanism of anisotropy from that in the ordered phase. 
Below we show that this significant anisotropy of specific heat can be consistently explained by the angle-dependent Majorana gap in the Kitaev QSL (see Fig.\,\ref{Hdiagram}b). 

\begin{figure}[h]
    \centering
    \includegraphics[width=0.8\linewidth,pagebox=artbox]{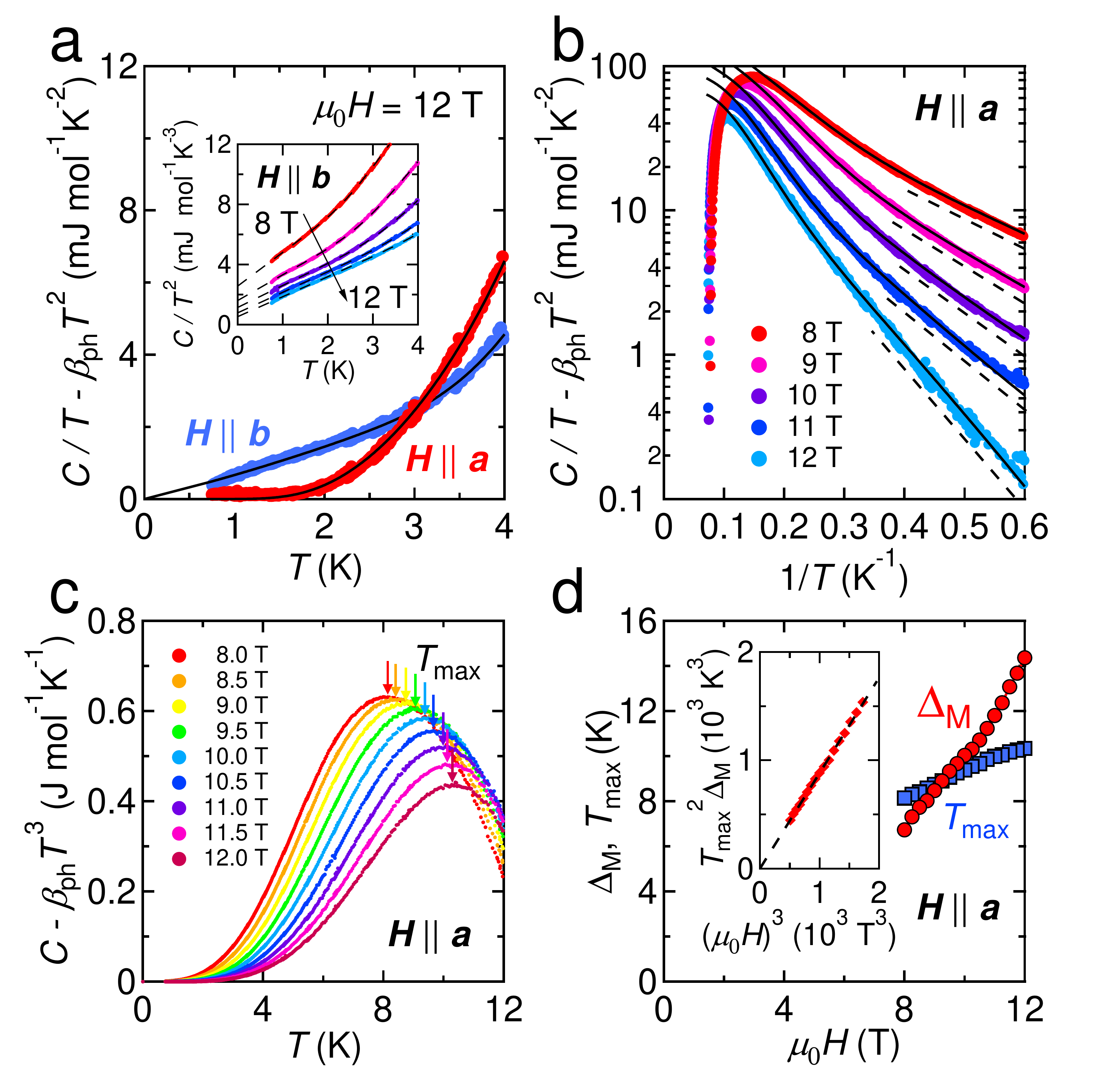}
    \caption{{\bf Temperature dependence of specific heat and field dependence of Majorana gap.}
    {\bf a}, Temperature dependence of $C/T-\beta_{\rm ph} T^2$ with $\beta_{\rm ph}=1.22$\,mJ mol$^{-1}$K$^{-4}$ for $\bm{H} \parallel \bm{a}$ (red) and for $\bm{H} \parallel \bm{b}$ (blue) at 12\,T. Lines are the fits to Eq.\,(\ref{fits}) with a gap $\Delta_{\rm M}=14.4$\,K for $\bm{H} \parallel \bm{a}$ and $\Delta_{\rm M}=0$ for $\bm{H} \parallel \bm{b}$.
    Inset shows the temperature dependence of $C/T^2$  for $\bm{H} \parallel \bm{b}$ between 8 and 12\,T with 1\,T intervals. Dashed lines are the fits to Eq.\,(\ref{fits}), and the extrapolated intersection and slope to $T\to 0$ yield $\alpha$ and $\beta$, respectively (see Fig.\,S2).
    {\bf b}, Arrhenius plot of $C/T-\beta_{\rm ph}T^2$ for $\bm{H} \parallel \bm{a}$ at several fields. Solid lines are the fits to Eq.\,(\ref{fits}). Dashed linear lines represent exponential temperature dependence. 
    {\bf c}, Temperature dependence of $C-\beta_{\rm ph} T^3$ for $\bm{H} \parallel \bm{a}$ at several fields. The peak temperature $T_{\rm max}$ (arrow) is related to the $Z_2$ flux gap $\Delta_{\rm flux}$. 
    {\bf d}, Field dependence of Majorana gap $\Delta_{\rm M}$ (red circles) and $T_{\rm max}$ (blue squares) for $\bm{H} \parallel \bm{a}$. Inset shows $T_{\rm max}^2\Delta_{\rm M}$ as a function of $(\mu_0 H)^3$. Dashed line represents $H^3$ dependence.
    }
    \label{Ctemp}   
\end{figure}

In the inset of Fig.\,\ref{Ctemp}a, we plot $C/T^2$ versus $T$ in the high-field paramagnetic phase for $\bm{H} \parallel \bm{b}$, which clearly indicates that specific heat has $\alpha T^2$ and $\beta T^3$ terms at low temperatures. 
The $\beta T^3$ term, which can be accurately determined from the low-temperature slope in this plot, corresponds to the bosonic contributions.
The obtained $\beta$ is field dependent, and approaches asymptotically to a constant value $\beta_{\rm ph}=1.22$\,mJ mol$^{-1}$K$^{-4}$ at high fields (Fig.\,S2c). 
This value is in good agreement with the estimate of phonon contribution from the comparison with isostructural nonmagnetic RhCl$_3$~\cite{Widmann2019}, where it has been shown together with {\it ab initio} calculations that the phonon gives strictly $\beta T^3$ term at low temperatures with $\beta\sim1.3$\,mJ mol$^{-1}$K$^{-4}$. 
The deviations of $\beta$ value from $\beta_{\rm ph}$ at lower fields imply the presence of additional bosonic contributions, likely related to antiferromagnetic fluctuations near the critical field (see Supplementary Information), but these contributions are not essential for the fermionic behaviors we focus below. 
By subtracting the $\beta_{\rm ph} T^2$ phonon term from the $C/T$ data, we find a crucial difference in the temperature dependence between the two field orientations, parallel and perpendicular to the Ru-Ru bond directions. 
As shown in Fig.\,\ref{Ctemp}a, $C/T-\beta_{\rm ph} T^2$ for $\bm{H} \parallel \bm{b}$ clearly shows a $T$-linear behavior at low temperatures, which corresponds to the finite residual $C/T^2$ in the zero-temperature limit (Fig.\,\ref{Ctemp}a, inset), demonstrating the presence of gapless excitations consistent with relativistic (massless) linear dispersions. 
In contrast, the data for $\bm{H} \parallel \bm{a}$ show exponential temperature dependence, which can be seen more clearly in the Arrhenius plot in Fig.\,\ref{Ctemp}b, indicating fully gapped excitations. 
Such strongly angle-dependent low-energy excitations in the paramagnetic phase, especially the gapless excitations for $\bm{H} \parallel \bm{b}$, cannot be easily explained by the conventional spin-wave-like excitations, but can be naturally accounted for by the angle-dependent Majorana gap in the Kitaev model (Fig.\,\ref{Hdiagram}b). 

To make further quantitative analysis of $C(T,H,\phi)/T$ in the paramagnetic state, we use the following formula, 
\begin{equation}
    C(T,H,\phi)/T=\beta(H,\phi) T^2 + C_{\rm M}(T,H,\phi)/T + C_{\rm flux}(T,H)/T.
    \label{fits}
\end{equation}
Here we consider bosonic contributions in the first $\beta T^2$ term, and the contributions from itinerant Majorana fermions ($C_{\rm M}/T$) and $Z_2$ flux ($C_{\rm flux}/T$) in the second and third terms, respectively.
The observed contrasting behaviors for the two directions can be quantitatively captured by the second $C_{\rm M}/T$ term in Eq.\,(\ref{fits}), i.e.\ the contribution from itinerant Majorana fermions with angle-dependent Majorana gap $\Delta_{\rm M}(\phi)$. 
This term can be given by the following formula obtained for a two-dimensional system with energy dispersion $E(k)=\sqrt{v^2k^2+\Delta_{\rm M}^2}$ (see Supplementary Information);
\begin{equation}
    C_{\rm M}(T;\Delta_{\rm M})/T =\alpha T\left(1-\frac{\mathcal{G}(\Delta_{\rm M}/T)}{\mathcal{G}(\infty)}\right), \quad     
    \mathcal{G}(y)\equiv  \int_0^y \frac{{\rm d}x}{2\pi} \frac{x^3{\rm e}^x}{({\rm e}^x+1)^2},
\label{Ceqn} 
\end{equation}
where $\alpha\propto 1/v^2$ is a factor determined by the velocity $v$ of Majorana fermions. 
The third $C_{\rm flux}/T$ term in Eq.\,(\ref{fits}) corresponds to the contribution from the $Z_2$ flux, which has a sizable gap, and is thus more sensitive to the high-temperature behavior.    
For this, we simply use the Schottky-type formula that mimics the two level system with a gap $\Delta_{\rm flux}$; 
$C_{\rm flux}/T\propto (\Delta_{\rm flux}^2/T^3) \cdot {\rm e}^{\Delta_{\rm flux}/T}/(1+{\rm e}^{\Delta_{\rm flux}/T})^2$.
The Majorana term Eq.\,(\ref{Ceqn}) gives a $T$-linear behavior of $C/T=\alpha T$ for gapless Majorana excitations (Fig.\,S3a), in agreement with the $\bm{H} \parallel \bm{b}$ data. 
For $\bm{H} \parallel \bm{a}$, the fitting with parameters $\alpha$ and $\beta$ shown in Fig.\,S2 yields a systematic evolution of gap magnitude $\Delta_{\rm M}$ as a function of field as shown in Fig.\,\ref{Ctemp}d. 
This increasing trend of $\Delta_{\rm M}$ with field is in qualitative agreement with the previous reports~\cite{Sears2017,Wolter2017}. 
This is also consistent with the change in $C(\phi)/T$ in Fig.\,\ref{Cangle}b from sinusoidal-like to more flat-bottom shape with increasing field, which can be accounted for by $C_{\rm M}(\phi)/T$ in Eq.\,(\ref{Ceqn}) with an increase of $\Delta_{\rm M}/T$ (see Fig.\,S3b).  
We emphasize that the observations of gapless excitations in one field direction and gapped behavior in another direction are quite striking, and low-energy excitations with such characteristic angle dependence are a hallmark of the Kitaev QSL. 

Next, to track the field dependence of $\Delta_{\rm flux}$, we can use the relation that the excitation gap of $Z_2$ flux is intimately related to the peak temperature $T_{\rm max}$ in the specific heat $C(T)$ of Kitaev QSL. 
It has been reported~\cite{Do2017,Widmann2019} that the magnetic contribution of $C(T)$ at high temperatures has a characteristic structure with two broad peaks at $\sim 10$\,K and $\sim 100$\,K, and the former is ascribed to the excitations of $Z_2$ flux~\cite{Motome2020}. 
The temperature dependence of $C-\beta_{\rm ph}T^3$ plotted in Fig.\,\ref{Ctemp}c clearly shows a peak structure at $T_{\rm max}$, which shows appreciable field dependence (Fig.\,\ref{Ctemp}d). 
This $Z_2$ flux peak is also linked to the Kitaev interaction $J_{\rm K}$, and quantum Monte Carlo simulations for the isotropic pure Kitaev model have shown the relations $T_{\rm max}\simeq 0.012 J_{\rm K}$ and $\Delta_{\rm flux}\sim0.07 J_{\rm K}$~\cite{Motome2020}.
This implies $T_{\rm max}\sim 0.17\Delta_{\rm flux}$, although quantitative estimates of $J_{\rm K}$ and $\Delta_{\rm flux}$ require realistic models of $\alpha$-RuCl$_3$ including non-Kitaev interactions.  
It has been also pointed out that the magnetic field can modify the Kitaev interaction~\cite{Nasu2018}.
Thus we infer that the field dependence of $T_{\rm max}$ accompanies the field dependence of $\Delta_{\rm flux}$. 
The simple relation $T_{\rm max}(H)\propto \Delta_{\rm flux}(H)$ enables us to check the validity of Eq.\,(\ref{Deq}) by plotting $T_{\rm max}^2\Delta_{\rm M}(H)$, which should be proportional to $|h_xh_yh_z|$. 
The obtained results shown in the inset of Fig.\,\ref{Ctemp}d demonstrate a distinct $H^3$ behavior, in excellent agreement with the Kitaev model. 

\begin{figure}[h]
    \centering
    \includegraphics[width=\linewidth,pagebox=artbox]{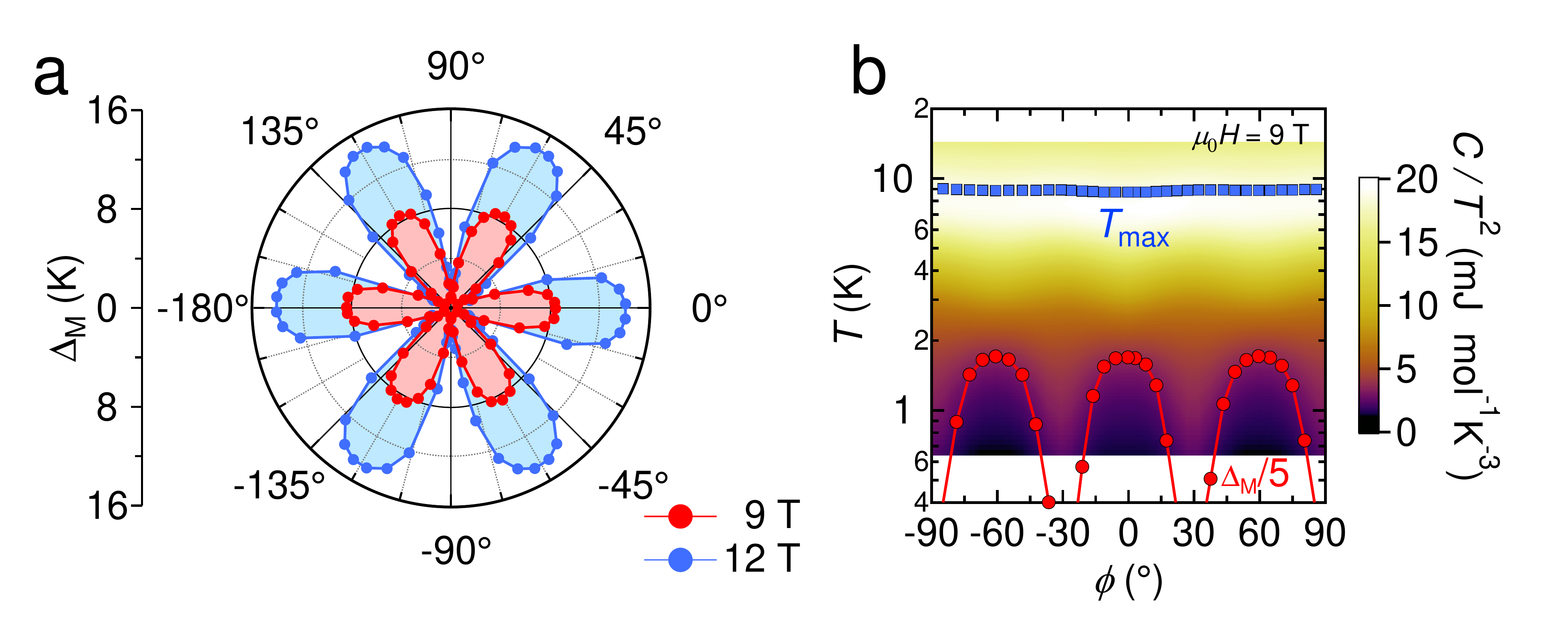}
    \caption{{\bf Field-angle dependence of Majorana gap.}
    {\bf a}, Polar plot for angle-dependent Majorana gap magnitude $\Delta_{\rm M}(\phi)$ for 9 and 12\,T. The data for $-90^\circ \le \phi \le 90^\circ$ are determined from the fits to the temperature dependence of $C/T$. We use the symmetrization to represent the other (left half) part by rotating the right half data by $180^\circ$, which corresponds to time reversal. 
    {\bf b}, Angle dependence of the Majorana gap $\Delta_{\rm M}$ (with a multiplication of $1/5$) and $T_{\rm max}$ (which is related to the $Z_2$ flux gap $\Delta_{\rm flux}$) at 9\,T. Superimposed color map represents the magnitude of $C/T^2$ at 9\,T. 
    }
    \label{Dangle}   
\end{figure}

Having established the field dependence of Majorana gap, we now turn to the angle dependence of $\Delta_{\rm M}$. 
We repeat the same fitting procedures for different angles at 9 and 12\,T, and the obtained $\Delta_{\rm M}(\phi)$ data are demonstrated in a polar plot of Fig.\,\ref{Dangle}a.
This captures the essential features of angle-dependent Majorana gap expected in the Kitaev QSL. 
First, the thermodynamically determined excitation gap shows a strong anisotropy in the honeycomb plane, having sextuple nodes along the $b$ axis and equivalent directions. 
Second, the angle dependence can be approximated by $|\cos3\phi|$, consistent with the $f$-wave symmetry of the angle dependence of the gap.
This strong angle dependence can also be visually seen in the color plot of $C(\phi)/T^2$ in Fig.\,\ref{Dangle}b, where the large gap corresponds to the suppressed specific heat at low temperatures. 
We note that at high temperatures the peak temperature $T_{\rm max}$ is essentially angle independent, which implies that the $Z_2$ flux gap $\Delta_{\rm flux}$ and hence the Kitaev interaction $J_{\rm K}$ do not depend on field angle although these depend on field magnitude as discussed above. 
This ensures that the observed $f$-wave Majorana gap with six nodes in the plane is fully consistent with Eq.\,(\ref{Deq}), providing further compelling evidence of the Kitaev QSL state with characteristic Majorana fermion excitations in this material. 
It has been theoretically shown that the presence of the non-Kitaev interactions can modify the field-angle dependence of Majorana gap significantly, but the sextuple node structure within the plane is robust against these non-Kitaev terms~\cite{Go2020}. 
Our results thus indicate that the fundamental property of Kitaev QSL having strong field-angle dependence with gapless Majorana excitations for Ru-Ru bond directions is preserved in the high-field paramagnetic phase of $\alpha$-RuCl$_3$.

We point out that recent thermal Hall conductivity measurements have shown that the half-integer quantized plateau behavior is observed for $\bm{H} \parallel \bm{a}$, but is absent for $\bm{H} \parallel \bm{b}$~\cite{Yokoi2020preprint}. 
This is fully compatible with our angle dependence of Majorana gap $\Delta_{\rm M}(\phi)$, which is directly related to the Chern number $Ch$ that vanishes for $\bm{H} \parallel \bm{b}$. 
These results thus demonstrate the bulk-edge correspondence in this QSL state; the Majorana edge mode, which is manifested by the half-integer quantum thermal Hall effect, is protected by the gapped excitations in the bulk, despite the presence of non-Kitaev interactions. 

The obtained angular variations of $C/T$ also provide an important information on the possible field-induced topological quantum phase transition in the high-field paramagnetic state suggested by some experiments~\cite{Kasahara2018,Yokoi2020preprint,Balz2019}. 
The thermal Hall conductivity deviates from the half-integer quantum value and vanishes above a characteristic field, whose in-plane component varies from $\sim 8$ to $\sim 11$\,T depending on samples and angles in the $ac$ plane~\cite{Kasahara2018,Yokoi2020preprint,Yamashita2020}. 
Neutron scattering and magnetocaloric effect measurements also point to some change around $\sim 9$\,T at 1.5\,K~\cite{Balz2019}. 
In the present specific heat study, the field dependence of the Majorana gap (Fig.\,\ref{Ctemp}d) continues to increase above $\sim 10$\,T, implying the absence of a certain continuous quantum phase transition that requires the closing of the gap~\cite{Go2019}. 

However, a close look at the field dependence of $C/T$ for $\bm{H} \parallel \bm{a}$ (Fig.\,\ref{Hdiagram}e) reveals a small bump at $\mu_0H^\ast\sim 10$\,T, which is not visible at $\phi=60^{\circ}$ (see Fig.\,S4b). 
Similarly, in the angle dependence of $C/T$ at 10\,T (Fig.\,\ref{Cangle}b) a small peak-like bump structure can be seen around $\phi=0^{\circ}$, which is absent near $\phi=60^{\circ}$. 
At higher fields, this bump structure shifts to positive and negative angles near $\phi=0^{\circ}$ and new bump structures appear at $\pm60^{\circ}$ and $\pm120^{\circ}$ (arrows in Fig.\,\ref{Cangle}b and Fig.\,S4c). 
The positions of these small anomalies are plotted in the $H$-$\phi$ diagram of Fig.\,\ref{Cangle}a, which indicates that this anomaly field $H^\ast$ has a strong two-fold ($C_2$) anisotropy in addition to the six-fold oscillations. 
One may notice a tiny $C_2$ anisotropy of $C(\phi)/T$ at lower fields (Fig.\,\ref{Cangle}c), which is likely related to the small structural distortion less than 0.2\% along the $b$ axis in $\alpha$-RuCl$_3$~\cite{Cao2016}.
However, the high-field $C_2$ anisotropy of $H^\ast(\phi)$ is much more significant ($\sim15$\%). 
As discussed in Supplementary Information, this significant $C_2$ anisotropy at high fields is not coming from the field misalignment and thus it has a different origin from the small structural distortion. 
We note that the effect of stacking faults, which are unavoidable in actual crystals of $\alpha$-RuCl$_3$, is known to lead to additional magnetic transitions at $\sim 10$-14\,K higher than $T_{\rm N}\approx7$\,K~\cite{Cao2016}. 
Although the detailed field dependence of these higher magnetic transitions is difficult to be tracked~\cite{Kubota2015}, the critical fields of these magnetic transitions are anticipated to be higher than 7\,T, and may have a two-fold angular anisotropy, considering the spin flop expected for the magnetic structure of the $C_2$-symmetric ABAB stacking faults within the ABCABC layer sequence in $\alpha$-RuCl$_3$. 
In our crystal, however, only tiny anomalies in $C/T$ at 10 and 13\,K can be seen (Fig.\,S4a), which indicates that the volume fraction of the stacking faults inside the crystal is less than $\sim2$\%. 
It is therefore unlikely that this tiny portion is totally responsible for the vanishing of quantum thermal Hall effect. 
From our analysis of angular dependence of Majorana gap $\Delta_{\rm M}(\phi)$ and $\alpha(\phi)$, which is related to the Majorana velocity, we obtain $\Delta_{\rm M}(60^\circ)/\Delta_{\rm M}(0^\circ)\approx1.04$ (1.01) and $\alpha(60^\circ)/\alpha(0^\circ)\approx0.95$ (1.02) at 12\,T above $H^\ast$ (9\,T below $H^\ast$).
This suggests that the low-energy properties of Majorana fermions break the rotational symmetry at high fields in the bulk. 

It has been pointed out by Kitaev that the magnetic field generates the interactions among neighboring four Majorana fermions~\cite{Kitaev2006}. 
When these interactions are strong, a bond-like order of Majorana fermions may be induced with keeping spin-fractionalization, which is expected to lower the six-fold symmetry down to $C_2$ symmetry in the QSL state. 
Moreover, once the order parameter of such a bond-like order becomes large, we could have a first-order transition from the B phase in the Kitaev model to an anisotropic A (toric code) phase~\cite{Takahashi2021}. 
This topological transition accompanies the annihilation of the two gapped cones~\cite{Motome2020}, resulting in the vanishment of Chern number without closing the Majorana gap, consistent with the vanishing of quantum thermal Hall effect. 
Such a field-induced transition with broken rotational symmetry can be called as a nematic transition. In electronic nematic states, in general, domains of different anisotropy orientation could be formed in the sample. In the present case, however, the stacking faults as well as tiny structural distortions can pin the director of the C2 nematic QSL state, which is otherwise difficult to be resolved by the field rotation experiments. 
This reminds us that a nematic order with spontaneous rotational symmetry breaking has also been found in quantum Hall systems near the 5/2 fractional quantum Hall state~\cite{Samkharadze2016}, which is believed to be a prototypical non-abelian topological phase characterized by the gapped state in the bulk with topologically protected gapless edge currents. 

To sum up, we have studied the specific heat in the Kitaev material $\alpha$-RuCl$_3$ in detail under magnetic fields rotated within the honeycomb plane. 
In the high-field paramagnetic phase, we have observed contrasting behaviors between gapped excitations for $\bm{H} \parallel \bm{a}$ and gapless excitations for $\bm{H} \parallel \bm{b}$, which are the features characteristic of itinerant Majorana fermions in the Kitaev QSL. 
The field dependence and angle dependence of the Majorana gap provide thermodynamic evidence that the Kitaev physics can be applied to the high-field state of $\alpha$-RuCl$_3$. 
We have also found a possible nematic transition in the Kitaev QSL state, which may be related to the vanishing of Chern number. 
Finally, our study demonstrates that the Majorana gap can be tuned by the field angle, which opens up a new pathway to control Majorana fermions in bulk materials.

\bibliographystyle{naturemag}
\bibliography{refNP}

\bigskip

\section*{Methods}

\subsection*{Single crystals}
High-quality single crystals of $\alpha$-RuCl$_3$ were grown by the vertical Bridgman method~\cite{Kubota2015}. To conduct heat capacity measurements, we carefully picked up the sample to minimize the stacking faults or bending of the crystal. The lateral size of the sample is $\simeq 1.1\times1.3$\,mm$^2$, and the weight is $\simeq 0.7$\,mg. To check the sample quality, we conducted the magnetization measurements and observed no discernible jump due to the additional magnetic transition originating from the stacking faults. The orientation of the crystal axis is determined by X-ray diffraction measurements. 

\subsection*{Heat capacity measurements}
Heat capacity of the crystal was measured by a long relaxation method which is designed for samples with small mass (see Supplementary Information). A bare chip resistive thermometer (Cernox 1030BR) is used as both the thermometer and the heater. The thermometer was calibrated under magnetic fields up to 14\,T by using a calibrated thermometer in a dilution fridge. The sample is directly attached on the bare chip using Apiezon grease for good thermal contact.

The holder is rotated by a piezo driven rotator equipped in a 12-T superconducting magnet. The angle of the chip is checked by the two Hall sensors orthogonally placed on the rotator. This whole system is mounted on a $^3$He cryostat with the base temperature of $\simeq 0.6$\,K. 

\newpage

\noindent
{\bf Acknowledgements}

\noindent
We thank A.~Go, C.~Hickey, K.~Hwang, Y.~Kasahara, M.~Matsuyama, Y.~Motome, J.~Nasu, M.~Udagawa, and M.~Yamashita for fruitful discussions. 
This work was supported by CREST (No.\ JPMJCR19T5) from Japan Science and Technology (JST), Grants-in-Aid for Scientific Research (KAKENHI) (Nos.\ JP19H00649, JP18H05227, JP20H02600, JP17H01142, JP19K03711, JP19K22123, JP18H01853, JP18KK0375), and Grant-in-Aid for Scientific Research on innovative areas ``Quantum Liquid Crystals" (No.\ JP19H05824) from Japan Society for the Promotion of Science (JSPS). E.-G.M. was supported by National Research Foundation of Korea under Grant NRF-2017R1C1B2009176,  NRF-2019M3E4A1080411, and NRF-2020R1A4A3079707.

\bigskip
\noindent
{\bf Author contributions}

\noindent
T.S. conceived and supervised the study. Y.Mizukami designed the measurement setup. O.T., Y.Mizukami, and R.H. performed the heat capacity measurements. O.T., Y.Mizukami, R.H., K.H., and T.S. analyzed the data with theoretical inputs from S.F. and E.-G.M. N.K. and H.T. synthesized the high-quality single crystalline samples. T.S. prepared the manuscript with inputs from Y.Mizukami, Y.Matsuda, and E.-G.M.

\bigskip
\noindent
{\bf Correspondence authors} 

\noindent
Correspondence to Yuta Mizukami or Takasada Shibauchi.

\bigskip
\noindent
{\bf Competing interests}

\noindent
The authors declare no competing interests.

\newpage

\end{document}